\documentclass[11pt]{article}

\usepackage{url}
\usepackage{hyperref}
\usepackage{graphicx}
\usepackage{amssymb}
\usepackage{amsmath}
\usepackage{amsthm}

\begin{document}

\begin{center}
\textbf{\begin{Large}Beyond the Fundamentals of Special Relativity:\end{Large}}
\end{center}

\begin{center}
\begin{Large}
\textbf{Full Lorentz $\gamma$ Factor}
\end{Large}
\end{center}

\begin{small}
\begin{center}\textbf{G. Sardin}\\Universitat de Barcelona - Facultat de Fisica\\gsardin@ub.edu
\end{center}
\end{small}

\begin{abstract}
\begin{small}

Special relativity calculates, by means of the Lorentz $\gamma$ factor, the proper time of all inertial systems from the observer proper time, which is taken as a time standard. So, any temporal inference relies in first instance on the observer own time. The question is thus: what fixes the observer proper time? This will be the crucial point debated here. This implies analyzing at the very first why the observer can be taken as a motionless reference in spite of being himself inertial. Is this just an approximation, and if so, up to what extent can it be applied? The framework of special relativity is compared to an amended form in which the fact of taking himself as a reference does not allow the observer to overlook its own kinetics. So, the debate stands on which of these two formulations of the $\gamma$ factor is the exact one:
\end{small}

\begin{center}
\begin{large}$\gamma = \frac{\sqrt{1-(\frac{v_{1}}{c})^2}}{\sqrt{1-(\frac{v_{1}+v}{c})^2}}$ \end{large} \qquad\qquad \begin{small}versus\end{small} \qquad\qquad \begin{large}$\gamma = \frac{1}{\sqrt{1-(\frac{v}{c})^2}}$ \end{large}
\end{center}

\begin{small}
The first formulation takes into account the fact that the observer is himself inertial, while the second disregards it. Already note that if the observer speed $v_{1}$ is ignored, the two formulations become identical. Hence, the standard relativistic expression of $\gamma$ can be seen as an approximation applicable when the observer motion is null or low, such as it is the instance on Earth.
\\
\textbf{Pacs:} 03.30.+p (Special Relativity), 45.20.D (Newtonian mechanics), 12.90.+b (Miscellaneous theoretical ideas and models)\\	
\end{small}
\end{abstract}

\section{Introduction}

Special relativity has so long fit all experimental data, so it has been considered ascertained.  Here, the operability of special relativity is therefore scrutinized. Special relativity takes the proper system as a sovereign referential frame, to which all inertial systems are referred to, thus allowing it to deal exclusively with speeds relative to the observer and to disregard the fact that himself is actuality inertial. Hence, let us scrutinize the support and workability of its approach [1-10]. It suffices to analyze why the actual motion of inertial observers can be allegedly disregarded, and up to what extent this applies. 
\\\\
Nowadays more experimental information is available than that Einstein had, and it is now known that our galaxy has a drift speed of about 600 km/s relative to the CMB [11-13], and that due to the galaxy rotation the Earth has a net speed of only about 370 km/s, which is relatively low compared to the speed of light. So, its speed is far away from that required for relativistic effects acting on it to be significant. But, even though its speed would be substantially higher special relativity would still work out. The motive is that, as long as the speed of the inertial observer through the CMB stands in the portion of the $\gamma$ curve in which it is almost linear, no significant quantitative departure from special relativity will come out. In that range of speed all inertial systems can be regarded equivalent. However, for an observer that would move at a relativistic speed through the CMB, and thus situated in the non linear range of the $\gamma$ curve, its own speed could not be neglected anymore, and special relativity would appear conceptually defective. Let us now demonstrate it. 
\\\\
It should be differentiated between classical observers, i.e. those with a proper speed in the linear lower range of the $\gamma$ curve and relativistic observers, i.e. those with a speed in the non linear part of the $\gamma$ curve. The standard $\gamma$ factor applies for classical observers as an acceptable approximation, but does not for relativistic observers in which case an amended full form of the $\gamma$ factor should be used.

\begin{center}\begin{LARGE}
$\gamma = \frac{\sqrt{1-(\frac{v_{1}}{c})^2}}{\sqrt{1-(\frac{v_{1}+v}{c})^2}}$ \end{LARGE} 
\qquad\qquad \begin{normalsize}versus\end{normalsize} \qquad\qquad \begin{LARGE}$\gamma = \frac{1}{\sqrt{1-(\frac{v}{c})^2}}$ 
\end{LARGE}\end{center}

The issue stands on which of these two formulations of the $\gamma$ factor is the accurate one. The first formulation of $\gamma$ takes into account the fact that the observer is himself inertial, while the second formulation disregards it. Let us note that if the observer speed $v_{1}$ is ignored the two formulations become identical. So, the standard expression of $\gamma$ can be regarded as being an approximation applying when the observer motion is null or low, such as for any observer on Earth. Let us now derive the amended formulation of $\gamma$ and the related proper time of inertial systems. 
\\\\
Special relativity foresees speed induced dilation of time, one of its upshots. Yet, one of its odd corollaries has been depicted in the Langevin or twin paradox [14-16]. The outcome of time dilation is here addressed not just from the standard mathematical handling but also from a physical approach concerned with causality. The point is that, since the proper time and time dilation are uncovered by means of clocks, let us center on their aptness to be sensitive to speed. 

\section{The proper-time issue}
Special relativity uses the observer proper time, and from it, defines the proper time of any inertial system with a relative speed $v$, according to the relation:
\\\\
(1) $t_{2} = \gamma$ $t_{1}$ = \begin{Large} $\frac{t_{1}}{\sqrt{1-\frac{v^{2}}{c^{2}}}}$ \end{Large} \quad and reciprocally: \quad (2) $t_{1}$ = \begin{Large}$\frac{t_{2}}{\gamma}$ \end{Large}= $t_{2}$ $\sqrt{1-\frac{v^{2}}{c^{2}}}$
\\\\
where $t_{1}$ is the observer proper time and $\gamma$ the Lorentz factor. The time $t_{2}$ so defined is thus the proper time of the observed inertial system. So, each inertial system has its own proper time and all them are correlated through their respective relative speed. In special relativity the observer takes himself as a sovereign reference, and defines the proper time of all inertial systems in base of his own, and through the exclusive reliance on the relative speed. So, a crucial question comes out, i.e. what fixes in first instance the observer proper time? Special relativity regards the proper time as a standard, disregarding any underlying dependence on the observer own motion through his spatial environment. 
\\\\
Obviously, the proper time of any inertial system cannot be causally ruled by the speed relative to any observer. So, any relation with the speed relative to an arbitrary observer has a relational value but certainly cannot have any causal assessment. The proper time of any inertial system must be ruled by some universal tenet, common to all of them. It surely cannot depend on any arbitrary referential observer that is himself inertial. Their proper time must be fixed by their speed relative to some universal extended entity, which in fact cannot be anything else but their own environment, i.e. space itself.  
\\\\
So, to reach understanding, from a causal viewpoint, the issue fixing the proper time of any inertial system, a universal reference is needed. From 1964, due to the fortuitous discovery by Arno Penzias and Robert Wilson [12] of the Cosmic Microwave Background (CMB), an extended reference is at hand. In effect, in view of its high isotropy [10-12] it constitutes a much valuable extended electromagnetic medium, furthermore of universal range. Moreover, the CMB being an electromagnetic field with an isotropic distribution implies that all photons embodying it go in all directions at the speed c within free space. Logically, this infers that the CMB cannot have any drift velocity through space. Consequently, the CMB or space itself can be referred equivalently in order to normalize speeds.
\\\\
Thus, let us define the proper time of any inertial system relative to the CMB, being so normalized to a reference common to all of them. The proper time $t_{1}$ of an observer on Earth expressed in relation with its speed $v_{1}$ relative to the motionless CMB, is:
\\\\
(3) $t_{1}$ = \begin{Large} $\frac{t_{0}}{\sqrt{1- \left(\frac{v_{1}}{c}\right)^{2}}}$ \end{Large} 
\\\\
Reciprocally, let us take our proper time $t_{1}$ on Earth as referential time, and from the speed $v_{1}$ in reference to the CMB let us deduce the proper time $t_{0}$ of a system at rest relative to it, i.e. at a lower speed of 370 km/s than that of the Earth:
\\\\
(4) $t_{0} = t_{1} \sqrt{1-\left(\frac{v_{1}}{c}\right)^{2}}$
\\\\
From now on we will refer all proper times with respect to $t_{0}$, allowing thus to normalize all them to a single and universal pattern of time measurement. So, the proper time $t_{2}$ of an inertial system with speed $v_{2}$ relative to the CMB is:
\\\\
(5) $t_{2} = t_{0}$ \begin{Large}$\frac{1}{\sqrt{1-\left(\frac{v_{2}}{c}\right)^2}}$ \end{Large} 
\\\\
Solving equations (4) and (5) gives:
\\\\
(6) $t_{2} = t_{1}$ \begin{Large}$\frac{\sqrt{1-\left(\frac{v_{1}}{c}\right)^2}}{\sqrt{1-\left(\frac{v_{2}}{c}\right)^2}}$ \end{Large} 
\\\\
but since $v_{2} = v_{1} + v$, ($v$ being the relative speed) we get:
\\\\	
(7) $t_{2} = t_{1}$ \begin{Large}$\frac{\sqrt{1-\left(\frac{v_{1}}{c}\right)^2}}{\sqrt{1-\left(\frac{v_{1}+v}{c}\right)^2}}$\end{Large} \: while the SR formula gives: \:(8) $t_{2}= t_{1}$ \begin{Large}$\frac{1}{\sqrt{1-\left(\frac{v}{c}\right)^2}}$\end{Large}	 	
\\\\
It merges out that the formulation of $\gamma$ from special relativity is short of the term  in the numerator and of the speed $v_{1}$ in the denominator, where $v_{1}$ refers to the actual speed of the proper inertial system through the CMB. It is easily seen that for low speed observers the two above formulations (7) and (8) tend to equality, and so $v_{1}$ can be neglected. However, this term becomes crucial and cannot be omitted anymore when the proper speed through the CMB, or equivalently through space, is high. So, it comes into sight that the formulation from special relativity is an approximation that can be used only when the speed of the proper inertial system through the CMB is low, such as it is the case for the Earth (Fig.1).
\\\\
\begin{figure}[h!]
\begin{center}
		\includegraphics[scale=0.6]{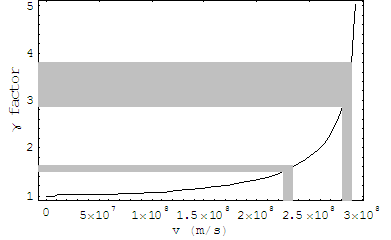}
\end{center}
\caption{\textit{It illustrates that for the same relative speed (width of the vertical strips) the $\gamma$ factor may have different values (width of the horizontal strips), according to the position of the relative speed along the $\gamma$ curve, i.e. according to the proper speed of the observer. Special relativity ignores this fact, giving always the same value of the $\gamma$ factor for the same relative speed. The different widths of $\gamma$ fixe the different proper-times of different observed inertial systems with the same speed relative to the observer.}}	
\end{figure}

\section{Influence of the position of the relative speed $v$ along the $\gamma$ curve}

Let us emphasize the need of taking into account the observer speed and its position along the $\gamma$ curve. Special relativity ignores in what range of the Lorentz $\gamma$ factor is situated the relative speed, which position depends on the observer proper speed relative to the CMB. This derives from overlooking the fact that the proper system is inertial, oversight which is a conceptual shortcut. In practice, it is not noticed due to the fact that the speed of Earth through space is low (370 km/s), so it can be ignored and Earth can be considered at rest. This allows all experiments made on it to be regarded as exclusively dependant on relative speeds from it. 
\\\\
\begin{figure}[h!]
\begin{center}
		\includegraphics[scale=0.6]{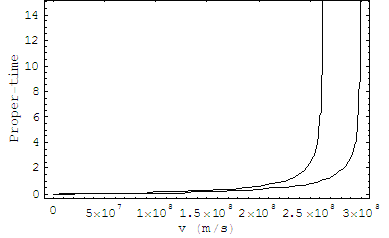}
\end{center}
\caption{\textit{Departure of the $\gamma$ factors according to the observer own speed. At low speed of the observer there is no significant difference between the two formulations of the $\gamma$ factor, but at high speed they depart considerably.}}	
\end{figure}
\\\\
However, a flaw would appear if the proper system would acquire a high speed through space. In that case its speed just could not be ignored, and the reduction to relative speeds would appear faulty, as already pointed out above. Special relativity stands safe thanks to the required high speed of the proper inertial system to evidence that it is conceptually defective is inaccessible. The reduction to relative speed works out for low speed observers only because in that range the Lorentz $\gamma$ factor is almost flat, so the influence of the proper speed is insignificant. However, if its speed would be high, above $10^{4}$ km/s, then it would be in the range of the $\gamma$ curve where it is no more almost linear, and hence the speed of the proper observer could no longer be neglected and the reduction to just relative speeds would fall short (Fig.2).

\section{Quantitative substantiations}

Let us evaluate three different cases with the same relative speed between observer and observed inertial object, e.g. of 30000 km/s, but with a different speed through space of the observer. This concept is unattended in special relativity, which only relies on the relative speed and thus it predicts the same value of the $\gamma$ factor for the three cases.

\subsection{The observer is at rest within space (or equivalently the CMB)}

It is clear that for $v_{1}$ = 0, equation (7) reduces to that of special relativity (8). So, the proper time $t_{2}$ of the observed object with a relative speed $v$ = 30000 km/s is in both formulations equal to:
\\\\
(8) $t_{2} = t_{1}$ \begin{Large}$\frac{\sqrt{1-(\frac{v_{1}}{c})^2}}{\sqrt{1-(\frac{v_{1}+v}{c})^2}} $\end{Large} = $t_{1}$ \begin{Large}$\frac{1}{\sqrt{1-(\frac{v}{c})^2}}$\end{Large} = 1.00504 $t_{1}$
\\\\ where $t_{1}$ is the observer proper time.
\\\\
It is therefore evidenced that when the observer speed is low it can be neglected and then $t_{2}$ is equal to that from special relativity. This makes evident why special relativity works out. So, special relativity is inadvertently assuming that the observer is at absolute rest.

\subsection{The observer is on Earth and so moving at the speed $v_{1}$ = 370 km/s in reference to the CMB}

For a proper speed $v_{1}$ = 370 km/s and the same relative speed $v$ = 30000 km/s as before:
\\\\
(9) $t_{2} = t_{1}$\:\begin{Large}$\frac{\sqrt{1-(\frac{v_{1}}{c})^2}}{\sqrt{1-(\frac{v_{1}+v}{c})^2}}$\end{Large} =\:1.00516 $t_{1}$ \quad and \quad $t_{2}= t_{1}$\:\begin{Large}$\frac{1}{\sqrt{1-(\frac{v}{c})^2}}$ \end{Large}=\:1.00504 $t_{1}$
\\\\
(10) $\Delta t$ = 1.00516 $t_{1}$ - 1.00504 $t_{1}$ = 0.00012 $t_{1}$
\\\\
It comes thus into view that when the proper motion is not disregarded the calculated  time $t_{2}$ differs from that of special relativity, which appears as an approximation valid only for low speed observers, such as it is the case on Earth.

\subsection{The observer is on an aircraft moving at a speed $v_{1}$ = 269.990 km/s in reference to the CMB}

For a proper speed $v_{1}$ = 269.990 km/s and the same relative speed $v$ = 30000 km/s as before:
\\\\
(11) $t_{2} = t_{1}$ \begin{Large}$\frac{\sqrt{1-(\frac{v_{1}}{c})^2}}{\sqrt{1-(\frac{v_{1}+v}{c})^2}}$ \end{Large} = 75.5046 $t_{1}$		and		$t_{2} = t_{1}$ \begin{Large}$\frac{1}{\sqrt{1-(\frac{v}{c})^2}}$ \end{Large} = 1.00504 $t_{1}$
\\\\
(12) $\Delta t$ = 75.5046 $t_{1}$ - 1.00504 $t_{1}$ = 74.4995 $t_{1}$
\\\\
Then, it appears clear that for a fast moving observer the time $t_{2}$ calculated from special relativity, which ignores the observer proper speed, would be substantially shorter. 

\subsection{The observed object is a spacecraft moving at a higher speed of 15 km/s than that of the Earth}

Since unfortunately we are unable to make the proper inertial system reaching very high speeds nor decelerating it down to a null absolute speed, let us consider another case, which is much less optimal that the previous pedagogic ones but that could be feasible. 
\\\\
For a relative speed $v$ of 15 km/s, special relativity predicts a time-dilation equal to:
\\\\
(13) $t_{2}= t_{1}$ \begin{Large}$\frac{1}{\sqrt{1-(\frac{v}{c})^2}}$\end{Large} = 1.00000000125 $t_{1}$ 
\\\\
While the full expression of $\gamma$ predicts a slightly different time-dilation:
\\\\
(14) $t_{2} = t_{1}$ \begin{Large}$\frac{\sqrt{1-(\frac{v_{1}}{c})^2}}{\sqrt{1-(\frac{v_{1}+v}{c})^2}}$ \end{Large} = 1.0000000629 $t_{1}$ 
\\\\
(15) $\Delta t$ = 6.167 $10^{-8}$ $t_{1}$
\\\\
Indeed a very small discrepancy, quite difficult to measure. If $t_{1}$ is taken equal to 1 second then $t_{2}$ expresses its dilation, and $\Delta t$ would then express the departure between the two predicted dilations of one second, i.e. 6.167 $10^{-8}$ s. Supposing a one year long roundtrip the discrepancy between the two predicted time dilations would be:
\\\\
(16) $\Delta t = 6.167\:10^{-8}\: t_{1} = (6.167\:10^{-8})$(3600 24 365) = 1.95 s
\\\\
This clearly highlights why special relativity stands safe in spite of being conceptually defective. It would be very difficult to feasibly measure such a small discrepancy knowing that time dilation would also be affected by residual microgravity and eventually by tiny temperature fluctuation. However, this does not justify not being concerned by the conceptual shortcomings of special relativity.

\section{Yearly variation of the proper time} 

Due to the rotation of the Earth around the Sun the net speed of Earth varies along the year, according to the change of the orientation of the speed $v$ relative to its drift speed  $v_{1}$ through the CMB, the maximum and minimum speed reached being at the most: $v_{1} \pm v  =  370 \pm 30$ km/s. Taking as unit of time $t_{1}$ = 1 s when $v$ is orthogonal to $v_{1}$, the half-yearly maximum and minimum dilation of one second would be:
\\\\
(17) $t_{2} = t_{1}$ \begin{Large}$\frac{\sqrt{1-\left(\frac{v_{1}}{c}\right)^{2}}}{\sqrt{1-\left(\frac{v_{1}+v}{c}\right)^{2}}}$ \end{Large} = 1.0000001283 s
\\\\
(18) $t_{2} = t_{1}$ \begin{Large}$\frac{\sqrt{1-\left(\frac{v_{1}}{c}\right)^{2}}}{\sqrt{1-\left(\frac{v_{1}-v}{c}\right)^{2}}}$ \end{Large} = 0.9999998817 s
\\\\
(19) $\Delta t$ = 1.0000001283 - 0.9999998817 = 2.4667 $10^{-7}$ s $\approx$ 0.25 $\mu s$
\\\\
So, time does not flow evenly all along the year. However, the net effect is almost null, since its increase during a semester is almost totally compensated by its decrease during the next semester, with a residual yearly shift of the order of $10^{-7}$ s. In the same way that we are not perceptive of our own speed, the yearly fluctuation of time cannot be detected on Earth, but an observer at absolute rest could uncover it.

\section{Application to the twin paradox}

To further illustrate, by way of the well known twin paradox, the departure from special relativity, which deals with a single value of the $\gamma$ factor for a given relative speed, let us calculate the aging difference for e.g. a one year journey, for the three first cases considered in section (4), i.e. with different speeds of the observer but the same relative speed in all cases. 

\subsection{Time-dilation predicted by special relativity}

According to special relativity, the aging difference between the two twins is exclusively fixed by the aircraft speed relative to the Earth. The Earth is taken as a sovereign reference and its actual motion is ignored. So, e.g. for a one year journey at a relative speed of 30000 km/s, special relativity predicts that the time-dilation difference between the twins would be:
\\\\
(20) $t_{2}= t_{1}$ \begin{Large}$\frac{1}{\sqrt{1-(\frac{v}{c})^2}}$ \end{Large}- $t_{1}$ = (1.00504 - 1) $t_{1}$ = 0.005038 $t_{1}$
\\\\
Let us now foresee the twins aging difference according to the three cases in which the relative speed between twins is maintained fix but the speed of the referential twin is varying.

\subsection{Time-dilation, the reference twin being at rest relative to the CMB}

If the normalized speed $v_{1}$, i.e. that relative to the CMB is null, then: 
\\\\
(21) $t_{2} = t_{1}$ \begin{Large}$\frac{\sqrt{1-(\frac{v_{1}}{c})^2}}{\sqrt{1-(\frac{v_{1}+v}{c})^2}}$ \end{Large}- $t_{1} = t_{1}$ \begin{Large}$\frac{1}{\sqrt{1-(\frac{v}{c})^2}}$ \end{Large}- $t_{1}$
\\\\
For a relative speed $v_{1}$ between twins of 30000 km/s, the time-dilation difference between them is identical from both formulations and equal to:
\\\\
(22) $t_{2}$ = (1.005038 - 1) $t_{1}$ = 0.005038 $t_{1}$
\\\\
So, in this case the two formulations are equivalent and thus the prediction from special relativity can be seen as an approximation in which the speed of the referential system is ignored and thus taken for null. 

\subsection{Time-dilation difference predicted by the amended $\gamma$ factor, the reference twin being on Earth}

Since from the normalized reference (CMB) the observer proper speed is accounted for, the predicted time-dilation difference between the twins is:
\\\\
(23) $t_{2} = t_{1}$ \begin{Large}$\frac{\sqrt{1-(\frac{v_{1}}{c})^2}}{\sqrt{1-(\frac{v_{1}+v}{c})^2}}$ \end{Large}- $t_{1}$ = 1.005163 - 1 = 0.005163 $t_{1}$
\\\\
where $v_{1}$ is the speed relative to the CMB and $t_{1}$ the proper-time on Earth.
\\\\
For $v_{1}$ = 370 km/s and $v$ = 30000 km/s, the discrepancy with the time-dilation predicted by special relativity  (equ.22), which foresees a single issue for a given relative speed, ignoring the motion of the referential inertial system, is thus: 
\\\\
(24) $\Delta t$ = 0.005163 $t_{1}$ - 0.005038 $t_{1}$ = 0.000125 $t_{1}$
\\\\
For a journey of one year, the discrepancy between the two predicted aging of the traveling twin would be:
\\\\
(25) $\Delta t$ = 0.000125 $t_{1}$ = 0.000125 (60 24 365) = 65.7 min

\subsection{Time-dilation difference, the reference twin being moving at a speed $v_{1}$ = 269990 km/s and the other one at a speed  $v_{2}$ = $v_{1} + v$, with $v$ = 30000 km/s}

(26) $t_{2} = t_{1}$ \begin{Large}$\frac{\sqrt{1-\left(\frac{v_{1}}{c}\right)^{2}}}{\sqrt{1-\left(\frac{v_{1}+v}{c}\right)^{2}}}$ \end{Large}- $t_{1}$  = 74.5046 $t_{1}$
\\\\
So, the difference between equ.(26) and equ.(22) is:
\\\\
(27) $\Delta t$ = 74.5046 $t_{1}$ - 0.005038 $t_{1}$ = 74.4995 $t_{1}$
\\\\
where $t_{1}$ is the proper time of the reference twin traveling at a speed $v_{1}$ = 269990 km/s in reference to the CMB.
\\\\
For one year journey of the twin traveling at the speed $v_{1}$, from equ.(27) the discrepancy between the two predicted time-dilations is:
\\\\
(28) $\Delta t$ = 74.4995 $t_{1}$ = 74.4995 years

\subsection{Actual issue of the traveling twin: physical effect induced by high-speed}

From a physical stance, the flow of time is indicated by the beating of clocks, whose oscillatory frequency is affected by speed, slowing down at increased speed. Naturally, this applies to any kind of oscillators, and thus to the oscillatory frequency of atoms, molecules and organic molecules. The time-dilation should not be treated independently of oscillators and approached in an abstract way, disjoint from its actual contextual origin. It should always be kept in mind that the actual cause of the speed-induced time-dilation is the slowing down of the frequency of clocks, and more generally of any type of oscillator. 
\\\\
So, the physical approach to the effects of speed predicts that all molecules from any organic body will decrease their oscillation frequency with increased speed. This has a crucial consequence. The slowing down of the oscillatory frequency of biological molecules can affect adversely the metabolic activity, endangering the upholding of life. So, actually the twin travelling at a relativistic speed may fall sick due to the slowing down of its metabolism or even die, without understanding why due to a counterfeit epicentral viewpoint, which by taking himself as reference, makes him illusively think that the other bodies are moving relative to him. This is a quite different issue than that predicted by the theory of relativity, which foresees a happy end for the travelling twin, returning younger than his brother instead of sick or even dead. 
\\\\
It has thus been pointed out that the observer proper speed and the position of the relative speed along the $\gamma$ curve cannot be disregarded for high speeds. Moreover, each twin ages according to his own absolute speed, and their speed difference defines their aging variance. So, there is no ambiguous shortcoming needing to be dissipated, as in special relativity, about who ages slower due to the symmetry between the twins relative speed, arising from the velocity reciprocity principle that stipulates: "If the velocity of an inertial frame S' relative to another such frame S is v, then the velocity of S relative to S' is -v". In the amended $\gamma$ framework there is no symmetry between twins, all speeds being expressed with respect to a single reference. So, the twin who has aged less is unequivocally the one who has traveled at the higher absolute speed, or that is to say, at higher speed in reference to the CMB. 

\section{Conclusions}

Let us highlight that equivalence between the proposed formulation of the $\gamma$ factor and the corresponding one from special relativity is fulfilled only when the observer is at absolute rest. When it does not their quantitative predictions differ. This points out that special relativity is inadvertently assuming that the observer is at absolute rest, despite it repudiates this concept. Since the Earth speed through space is quite low, the departure of the standard $\gamma$ factor from its amended homologue is so small that it stays short of quantitative significance. Yet, on conceptual grounds the difference is essential, since it establishes that the laws of physics are determined by the absolute speed of the inertial systems. The curtailed reliance on relative speeds is applicable only for slow moving observers. On practice this will not affect the use of special relativity because we live on a slow moving habitat. 
\\\\
A genuine shortcoming comes from treating time just mathematically as if it were a sovereign entity disjointed from clocks. The dependence of time on speed derives from the reliance on speed of the beating of clocks. It also merges out that the speed-induced dilation of time cannot be thought as just applying on aging, without considering its full physical foundation. In effect, the flow of time is evidenced through the beating of clocks, which oscillatory frequency is affected by speed, slowing down as it increases. Since this applies to any kind of oscillators and thus to the oscillatory frequency of atoms, molecules and organic molecules as well, this would affect the metabolism of any space traveler.
\\\\
It has been evidenced that the $\gamma$ factor used by special relativity is a truncated formulation having the value of an approximation, suitable only for slow moving inertial observers. The reduction to relative speed works out for low speed proper systems such as the Earth only because in that range the Lorentz $\gamma$ factor is almost flat, so the influence of the proper speed is insignificant. However, on an aircraft with a speed above $10^{4}$ km/s, it would then be in the $\gamma$ factor range where its slope starts varying significantly and the omission of its own speed would not stand anymore. It urges recovering pragmatic foundations taking into account the decisive fact that the proper system is actually moving within space and that its speed cannot be neglected when it is high. The fact that physical laws come out to be identical, as an acceptable approximation for slow inertial observers, does not uphold setting a conceptual framework ignoring their actual motion through space. 
\\\\
It would be wishful that an agency with cutting-edge technological means would check the fundamental issue of the influence of the observer proper kinetics, issue discarded in the foundation of special relativity.

\section{Bibliography}
\begin{footnotesize}

\quad \:\: [1] G. Sardin, Full Nexus between Newtonian and Relativistic Mechanics (2008), \\
\url{http://arxiv.org/ftp/arxiv/papers/0806/0806.0171.pdf}

[2] G. Sardin, First and second order electromagnetic equivalency of inertial systems, based on the wavelength and the period as speed-dependant units of length and time (2004), \url{http://arxiv.org/ftp/physics/papers/0401/0401092.pdf}

[3] G. Sardin, Testing Lorentz symmetry of special relativity by means of the Virgo or Ligo set up, through the differential measure of the two orthogonal beams time-of-flight (2004), \url{http://arxiv.org/ftp/physics/papers/0404/0404116.pdf}

[4] Anders Mansson, Understanding the special theory of relativity (2009),\\ 
\url{http://arxiv.org/abs/0901.4690}

[5] D. Bohm, The Special Theory of Relativity, Ed. W.A. Benjamin, Inc., N.Y. (1965)

[6] H. P. Roberton, Postulate versus Observation in the Special Theory of Relativity, Rev. Mod. Phys., 21 (1949) pp.378-382

[7] B.G. Sidharth, Looking beyond special relativity (2006),\\ 
\url{http://arxiv.org/abs/physics/0603189}

[8] J.M. Carmona, J.L. Cortes, Departures from special relativity beyond effective field theories (2006), \url{http://arxiv.org/abs/hep-th/0601023}

[9] A.E. Chubykalo, A. Espinoza, B. P. Kosyakov, Inertial frames of reference, space and time measurements, and physical principles of special relativity revisited (2009),\\
\url{http://arxiv.org/abs/0906.5219}

[10] Young-Sea Huang, An alternative to relativistic transformation of special relativity based on the first principles (2009),\\ \url{http://arxiv.org/abs/0812.5029}

[11] I. D. Karachentsev, D. A. Makarov,  The Galaxy Motion Relative to Nearby Galaxies and the Local Velocity Field, Astronomical Journal v.111, 2, p.794 (1996), \\
\url{http://adsabs.harvard.edu/full/1996AJ....111..794K}

[12] NASA, the CMB, \url{http://map.gsfc.nasa.gov/universe/bb_tests_cmb.html}

[13] Wikipedia, Cosmic Microwave Background Radiation,\\
\url{http://en.wikipedia.org/wiki/Cosmic_microwave_background_radiation}

[14] Wikipedia, Twin paradox, \url{http://en.wikipedia.org/wiki/Twin_paradox}

[15] Ying-Qiu Gu, Some Paradoxes in Special Relativity (2009),\\ 
\url{http://arxiv.org/abs/0902.2032}

[16] Y. Terletskii, Paradoxes in the Theory of Relativity, Plenium Press Ed., New York (1968)

\end{footnotesize}
\end{document}